\documentclass[a4paper,11pt,reqno]{amsart}
\usepackage{amssymb}
\usepackage{graphicx}
\numberwithin{equation}{section}
\swapnumbers
\theoremstyle{plain}
\newtheorem{thm}{Theorem}[section]
\newtheorem{lem}[thm]{Lemma}
\theoremstyle{definition}
\newtheorem*{Def*}{Definition}
\newtheorem*{rems*}{Remarks}
\newtheorem*{rem*}{Remark}
\providecommand{\D}[1]{\mathbb{#1}}

\providecommand{\R}[1]{\mathrm{#1}}
\providecommand{\abs}[1]{\lvert#1\rvert}
\providecommand{\accol}[1]{\lbrace#1\rbrace}
\providecommand{\croch}[1]{\lbrack#1\rbrack}
\providecommand{\norm}[1]{\lVert#1\rVert}

\newcommand{\dd}{\mathrm{d}}
\newcommand{\ee}{\mathrm{e}}

\DeclareMathOperator{\card}{\#}

\DeclareMathOperator{\diam}{diam}
\DeclareMathOperator{\dist}{dist}
\DeclareMathOperator{\expect}{\D{E}}
\DeclareMathOperator{\prob}{\D{P}}
\DeclareMathOperator{\supp}{supp}
\begin{document}
%-------------------------------------------------------------%
\title[Absence of continuous spectrum for certain random models]
       {Absence of continuous spectral types for certain
        nonstationary random models}
\author[A.~Boutet de Monvel]{Anne Boutet de Monvel$^\ast$}
\author[P.~Stollmann]{Peter Stollmann$^\dagger$}
\author[G.~Stolz]{Gunter Stolz$^\ddagger$}
\address{$^{\ast}$ IMJ, case 7012, Universit\'e Paris 7,
              2 place Jussieu, 75251 Paris, France}
\address{$^{\dagger}$ Fakult\"at f\"ur Mathematik,
          Technische Universit\"at, 09107 Chemnitz, Germany}
\address{$^{\ddagger}$ Department of Mathematics,
          University of Alabama at Birmingham, Birmingham, AL 35294, USA}
\dedicatory{In Memory of Robert M. Kauffman}
%-------------------------------------------------------------%
\date{}
%-------------------------------------------------------------%
\begin{abstract}
We consider continuum random Schr\"odinger operators of the type
$H_{\omega} = -\Delta + V_0 + V_{\omega}$ with a deterministic
background potential $V_0$. We establish criteria for the absence
of continuous and absolutely continuous spectrum, respectively,
outside the spectrum of $-\Delta +V_0$. The models we treat
include random surface potentials as well as sparse or slowly
decaying random potentials. In particular, we establish absence of
absolutely continuous surface spectrum for random potentials
supported near a one-dimensional surface (``random tube'') in
arbitrary dimension.
\end{abstract}
%-------------------------------------------------------------%
\maketitle
%-------------------------------------------------------------%
\section{Introduction}                             \label{s1}
In this article we are concerned with spectral properties of
certain nonstationary random models. This type of models has
attracted considerable interest as it allows to study a transition
from pure point to continuous spectrum. Here, we are mainly
concerned with the former phenomenon. We obtain our results by
essentially ``deterministic'' techniques from \cite{S2,McGS2,S1}
which gives us considerable flexibility. In particular, we are
able to avoid some of the  typical technical restrictions that
come with the usual multiscale analysis or fractional moments
proofs of localization. E.g., we can allow for perturbations of
changing sign and single site distributions without any
continuity. On the other hand, we need decaying randomness in the
sense that near infinity the random perturbation is not too
effective. That excludes identically distributed random parameters
in most cases. An important exception is our result on 1-D
``surfaces'' (rather tubes) in arbitrary dimensions.

The paper is organised in the following way: In Section \ref{s2}
we present the techniques we use, recalling the relevant notions
and results from  \cite{S2,McGS2,S1}; in fact we will need results
that are a little stronger than what is explicitly stated in the
above cited articles. The common flavour of these methods is that
they provide comparison criteria for the absence of continuous and
absolutely continuous spectra, respectively. These criteria are
formulated in the following way: We consider Schr\"odinger
operators with two potentials that differ only on a set that is
``small near infinity in a certain geometrical sense''. Then the
spectrum of the first operator has no absolutely continuous
component on the resolvent set of the second one. To exclude
continuous spectrum one needs a bit more complicated assertions
involving randomization.

In Section \ref{s3} we are dealing with sparse random potentials.
The framework we  introduce is fairly general and includes as
special cases the sparse random models considered in \cite{HuKi},
e.g.\ random scatterers are distributed quite arbitrarily in space
and the single site perturbations are assumed to be picked with
probabilities that tend to zero near infinity. Then, throughout
the resolvent set of the unperturbed operator there is no
absolutely continuous spectrum. Since we can treat quite general
unperturbed operators, this includes cases with gaps in the
spectrum of the unperturbed operator, a case that is completely
new. In the proof we combine elementary combinatorial arguments,
Lemma \ref{lem1}, with the methods discussed above. In the same
fashion, under a bit more incisive conditions concerning the
background and at least one random scatterer but with the same
condition concerning the decay of probabilities near infinity, we
can even deduce absence of continuous spectrum outside the
resolvent set of the unperturbed operator. This is quite different
from what one can obtain with the usual localization proofs. These
proofs require some disorder condition, or apply to energies near
the gaps only (with the exception of the one-dimensional case).

In Section 4 we study surface-like structures. This means we
consider potentials that are concentrated near a subset of lower
dimension. Our strongest result, Theorem \ref{thm.4.1} concerns
what we call quasi-1D surfaces. There is quite some literature on
surface potentials. Most are dealing with the discrete case
\cite{BS:98, c1, JL, JL2, JM:98, JM:99a, JM:99b, JMP} while in
\cite{bs,HuKi} and the present paper continuum models are treated.
Here again, our goal was to be able to exclude absolutely
continuous spectrum on all of the unperturbed resolvent set.

In the last section we conclude with a discussion of some possible
extensions of our results and a comparison with other works, in
particular the results in \cite{JM:99a} and \cite{HuKi}.\\

%-------------------------------------------------------------%
\section{Comparison criteria for absence
          of (absolutely) continuous spectrum}       \label{s2}

In this section we present our methods of proof, essentially
taken from \cite{S2,McGS2,S1}.
These methods rely on  comparison of the spectral properties of
Schr\"odinger operators
\[
H_1=-\Delta+V_1\mbox{  and  } H_2=-\Delta+V_2
\]
whose ``difference" is ``small" in the sense that the set
\[
\accol{V_1\neq V_2}:=\accol{x\in\D{R}^d\mid V_1(x)\neq V_2(x)}
\]
is sufficiently sparse.
To this end we introduce the following concept, following \cite{S2}:

%-----------%
\begin{Def*}
A sequence $(S_n)_{n\in\D{N}}$ of compact subsets of $\D{R}^d$
with Lebesgue measure $\abs{S_n}=0$ ($n\in\D{N}$) is called a
\emph{total decomposition} if there exists a family $(U_i)_{i\in
I}$ of disjoint, open, bounded sets such that
\[
\D{R}^d\setminus\bigcup_{n\in\D{N}}S_n=\bigcup_{i\in I}U_i.
\]
A typical example would be $S_n=\partial B(0,n)$, where $B(x,r)$
denotes the closed ball of radius $r$, centered at $x$. (Let us
stress that the $S_n$'s need not be pairwise disjoint.)
\end{Def*}
%-----------%

The sparseness of $\accol{V_1\neq V_2}$ will be expressed by the
existence of a
total decomposition $(S_n)_{n\in\D{N}}$ with sufficient distance of
$S_n$ to $\accol{V_1\neq V_2}$
compared with the \emph{size} of $S_n$.
An appropriate notion of size is given by the
\emph{generalized surface area} of a set, a notion introduced in
\cite{McGS2} in the following way; here $S\subset\D{R}^d$ is compact:
\[
\sigma(S):=
\sup_{r\geq 0}\frac{\abs{\accol{x\in\D{R}^d\mid
                     r\leq\dist(x,S)\leq r+1}}}{r^d+1}.
\]
It is easily seen that
\[
\sigma(S)\leq C\,((\diam S)^d+1),
\]
i.e.\ $\sigma(S)$ is at worst a volume, while for sufficiently
regular surfaces it is a surface area measure, for example
\[
\sigma(\partial B(x,r)) \le C(r^{d-1}+1).
\]

We cite the following result, essentially taken from \cite{S2}:

%-----------%
\begin{thm}                               \label{thm.ac}
Assume that for each $\gamma>0$ there exists a total decomposition
$(S_n)_{n\in\D{N}}=(S_n^{(\gamma)})_{n\in\D{N}}$ such that
\begin{equation}                          \label{eq.2.0}
\delta_n = \delta_n^{(\gamma)}:=\dist(\accol{V_1\neq V_2},S_n) \to
\infty \text{ as } n\to\infty
\end{equation}
and
\begin{equation}                            \label{eq.2.1}
\sum_n\sigma(S_n)\ee^{-\gamma\delta_n}<\infty.
\end{equation}
Then
\[
\sigma_{\textup{ac}}(H_1)\cap\varrho(H_2)=\varnothing.
\]
\end{thm}
%-----------%

The following figure is to help visualizing the geometry one is confronted 
with in the Theorem.

%-----------------%
\begin{figure}[h]
  \centering
        \includegraphics[width=8cm]{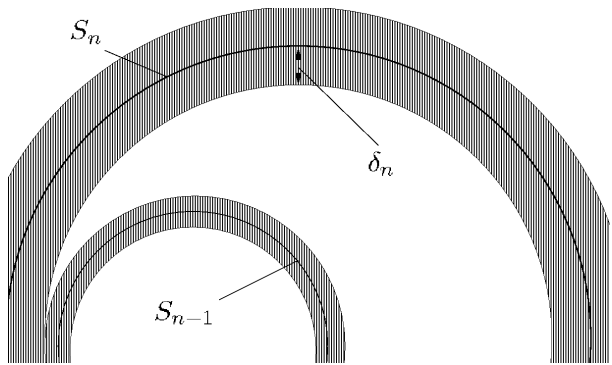}
        \caption{$\{ V_1\neq V_2\}$ must not intersect the shaded region}
        \label{decomposition}
\end{figure}
%------------------%
Here, and in what follows, all potentials $V$ are assumed to be
locally uniformly in $L^p$, where $p\ge 2$ if $d\le 3$ and $p>d/2$
if $d>3$, i.e.\
\begin{equation}                              \label{eq.2.1b}
\|V\|_{p,unif}^p := \sup_x \int_{B(x,1)} \abs{V(y)}^p\,dy <
\infty.
\end{equation}

Theorem~\ref{thm.ac} is essentially Theorem~4.2 from \cite{S2}. We
will need the slightly stronger version provided above in which

the decomposition $S_n$ may vary with $\gamma$. The proof provided
in \cite{S2} goes through under this weaker assumption. This is
roughly seen as follows: It suffices to show that
\begin{equation}                               \label{eq.2.1a}
\sigma_{\textup{ac}}(H_1) \cap J = \varnothing
\end{equation}
for all compact subsets $J$ of $\varrho(H_2)$. For fixed $J$ the
argument in \cite{S2} provides an $\gamma>0$ (roughly the
exponential decay rate in a Combes-Thomas type bound on the
resolvent of $H_2$ for energies in $J$) such that the validity of
(\ref{eq.2.0}) and (\ref{eq.2.1}) for a suitable decomposition
will imply (\ref{eq.2.1a}).

Also, in \cite{S2} all potentials are assumed to have locally
integrable positive parts and negative parts in the Kato class.
Our $L^p$-type assumptions are a special case.

The second result we use is taken from \cite{S1} and excludes
continuous spectrum. It is clear that a statement of the form of
Theorem 2.1 above has to be false, since dense pure point spectrum
is extremely instable and can be destroyed by ``tiny"
perturbations \cite{Simon:rank1}. The geometry is somewhat similar
to what we had above but more restrictive. Namely, consider an
increasing sequence $(A_n)_{n\in\D{N}}$ of bounded open sets with
$\bigcup_n A_n=\D{R}^d$. Then $S_n:=
\partial A_n$ is a total decomposition. For the arguments in
\cite{S1} it is not necessary that $\abs{\partial S_n} = 0$, but
this will be the case in all our applications.

We assume that
\[
\delta_{n}':=\min
\accol{\dist(S_n,\accol{V_1\neq
V_2}),\tfrac{1}{2}\dist(S_n,S_{n-1}\cup S_{n+1})}>0.
\]

%-----------%
\begin{thm}                               \label{thm.2.2}
Assume that $V_1\in L_{\textup{loc}}^{\frac{d+1}{2}}(\D{R}^d)$,
$W\in L^{\infty}$ with compact support, of fixed sign and such
that $|W| \ge c \chi_{B(0,s)}$ for suitable $c>0$ and $s>0$.
Moreover, assume that for every $\gamma>0$ there exist $A_n =
A_n(\gamma)$ as above such that $\delta_n'=\delta_n'(\gamma) \to
\infty$ and
\begin{equation}                            \label{eq.2.2}
\sum_n\abs{A_{n+1}\setminus A_{n-1}}\,\ee^{-\gamma
\delta_n'}<\infty.
\end{equation}
Then for the family $H_{\lambda}:= H_1+\lambda W$,
$\lambda\in\D{R}$ there exists a measurable subset
$M_0\subset\D{R}$ such that $\abs{\D{R}\setminus M_0}=0$ and
\[
\sigma_{\textup{c}}(H_{\lambda})\cap\varrho(H_2)=\varnothing
\text{ for all }\lambda\in M_0.
\]
\end{thm}
%-----------%

See \cite{S1} for the proof which extends to the case of $W$ as
specified above.

Again, as with Theorem~\ref{thm.ac} above, the possible
$\gamma$-dependence of the sets $A_n$ is not explicitly stated in
\cite{S1}, but allowed for by the proof provided there.

The requirement that the summability conditions \eqref{eq.2.1},
\eqref{eq.2.2} have to hold for \emph{all} $\gamma>0$ (and
suitable decompositions) come from the fact that we want to
exclude (absolutely) continuous spectrum up to the edges of
$\sigma(H_2)$. It is possible to quantify and refine the results
in a way which says that validity of \eqref{eq.2.1},
\eqref{eq.2.2} for a fixed $\gamma$ implies absence of
(absolutely) continuous spectrum in regions above a certain
($\gamma$-dependent) distance from $\sigma(H_2)$.

%-------------------------------------------------------------%
\section{Sparse random models}       \label{s3}

In this section we will show how to use the methods from the preceding
section to prove absence of continuous or absolutely continuous spectrum for
sparse random potentials. As mentioned in the introduction, these models have
been set up to study situations in which a transition from singular to
absolutely continuous spectrum occurs.

This has attracted some interest in the last decades as can be seen in the
articles \cite{ki,kikrio,kri1,kri2,krisi,m,mv} dealing with discrete
Schr\"odinger operators and \cite{HuKi} for the continuum case.

We will be concerned mainly with absence of a continuous spectral
component away from the spectrum of the unperturbed operator. For
this reason we state our results in a generality that does include
cases in which no absolutely continuous spectrum survives. As
model examples, let us mention two families of models that have
been treated in \cite{HuKi}.

Throughout, the single site potentials will be assumed to be
compactly supported and in $L^p$, $p$ as above.

In fact, it will be sufficiently interesting to think of compactly supported,
bounded $f$ as done in \cite{HuKi}. However, we shall not
restrict ourselves to a fixed sign of $f$.

Specific models of sparse random potentials, as considered in
\cite{HuKi}, are

%----------------------%
\subsubsection*{\bf Model 1}
\[
V_{\omega}(x)=\sum_{i\in\D{Z}^d}\xi_i(\omega)f(x-i),
\quad\omega\in\Omega
\]
where the $\xi_i$ are independent Bernoulli variables. Set
$p_i:=\prob(\xi_i=1)$. If $p_i\to 0$ as $\abs{i}\to\infty$ the
random potential will no longer be stationary. In fact, it will be
sparse in the sense that almost surely large islands near $\infty$
will occur where $V_{\omega}$ vanishes.

For the second model the $\xi_i$ and $p_i$ will have the same
meaning and, additionally, the $q_i$ are i.i.d.\ nonnegative
random variables.

%----------------------%
\subsubsection*{\bf Model 2}
\[
V_{\omega}(x)=\sum_{i\in\D{Z}^d}q_i(\omega)\xi_i(\omega)f(x-i).
\]
Again, $V_{\omega}$ is sparse in the above sense. Of course, for
$p_i\equiv 1$ we would get the
usual Anderson model.
Hundertmark and Kirsch study in \cite{HuKi}
the metal insulator
transition for $H(\omega)=- \Delta+V_{\omega}$ in $L^2(\D{R}^d)$ for the
case that $p_i\to 0$
as $\vert i\vert \to\infty$ but not too fast in order to
make sure that
$\sigma_{\text{ess}}(H(\omega))\cap(-\infty,0)\neq\varnothing$.

\medskip

In the following we consider:
\begin{itemize}
\item[(A$_1$)] $V_0:\D{R}^d \to \D{R}$ which is locally uniformly
$L^p$ with $p\ge 2$ if $d\le 3$ and $p>d/2$
if $d>3$.
\item[(A$_2$)] $\Sigma\subset\D{R}^d$ a set of sites that is
uniformly discrete in the sense that
\[
\inf\accol{\abs{j-i}\mid j,i\in\Sigma,\,j\neq i}=:r_{\Sigma}>0.
\]
\item[(A$_3$)] For each $i\in\Sigma$ a single site potential
$f_i\in L^p$ such that, for finite constants $\rho$ and $M$,
\[
\supp f_i\subset B(0,\varrho) \text{ and }\norm{f_i}_p\leq M.
\]
\item[(A$_4$)]
\[
V_{\omega}(x)=\sum_{i\in\Sigma}\omega_i f_i(x-i)
\]
where $\omega = (\omega_i)_{i\in\Sigma} \in (\Omega,\prob) =
(\D{R}^{\Sigma}, \bigotimes_{i\in\Sigma} \mu_i)$, i.e.\ the
$\omega_i$ are independent random variables with distribution
$\mu_i$, and $\supp \mu_i \subset [0,1]$ for all $i\in \Sigma$.
\end{itemize}

For our results on absence of continuous spectrum, in order to
apply Theorem~\ref{thm.2.2}, we will also require
\begin{itemize}
\item[(A$_5$)] Let $V_0$, $f_i \in
L_{\textup{loc}}^{(d+1)/2}(\D{R}^d)$ for all $i\in \Sigma$. There
exists one $k\in \Sigma$ with $f_k$ of definite sign, bounded, and
such that $|f_k| \ge c\chi_{B(0,s)}$ for some $c>0$ and $s>0$.
\end{itemize}

For further reference denote
\begin{equation}                        \label{eq.3.1}
p_i(\varepsilon):=
\mu_i(\croch{\varepsilon,1})=\prob\accol{\omega_i\geq\varepsilon}.
\end{equation}
Also, denote by
\begin{equation}                        \label{eq.3.1a}
m_k := (\mu_k)_\textup{ac}([0,1]),
\end{equation}
the total mass of the absolutely continuous component
$(\mu_k)_{\textup{ac}}$ of $\mu_k$. We will only use this for the
fixed $k\in\Sigma$ given in (A$_5$).

We consider the self-adjoint random Schr\"odinger operator
\begin{equation}                         \label{eq.3.2}
H(\omega)=H_0+V_{\omega}\text{ in }L^2(\D{R}^d)
\end{equation}
where $H_0 = -\Delta +V_0$. Our assumptions guarantee that the
local $L^p$-bounds (\ref{eq.2.1b}) for $V_0+V_{\omega}$ are
uniform not only in $x$, but also in $\omega$.

Of course, our model contains models I and II above as special
cases and $p_i(\varepsilon)\leq p_i$ for any $\varepsilon >0$ in
these cases. We have the following result:

%-----------%
\begin{thm}                               \label{thm.3.1}
Let $H(\omega)$ be as above, satisfying (A$_1$) to (A$_4$), and
assume that for all $\varepsilon
>0$,
\begin{equation}                        \label{eq.3.3}
p_i(\varepsilon) = o(\abs{i}^{-(d-1)}) \text{ as }
\abs{i}\to\infty.
\end{equation}
Then

\textup{(a)}
$\sigma_{\textup{ac}}(H(\omega)\cap\varrho(H_0)=\varnothing$
almost surely.

\textup{(b)} Assume, moreover, that (A$_5$) holds. Then, with $k$
as in (A$_5$),
\begin{equation}                                  \label{eq.3.4}
\prob\accol{\sigma_{\R{c}}(H(\omega))\cap\varrho(H_0)=\varnothing}\geq
m_k.
\end{equation}
\end{thm}
%-----------%

In particular,
$\sigma_{\R{c}}(H(\omega))\cap\varrho(H_0)=\varnothing$ holds
almost surely if $\mu_k$ is purely absolutely continuous, without
any assumption on the distribution at the other sites.

In order to apply the results from Section~\ref{s2} we need to
find sufficiently many and sufficiently large regions in which the
random potential $V_{\omega}$ is small and thus $H_{\omega}$ close
to $H_0$. We start by showing that these regions appear with
probability one.

Call a set $U$ $\varepsilon$-free for $\omega$ if $\omega_i \le
\varepsilon$ for all $i\in \Sigma \cap U$.

Denote by
\begin{equation}                          \label{eq.3.4a}
A_{r,R} = B(0,R) \setminus B(0,r)
\end{equation}
the annulus with
inner radius $r$ and outer radius $R$.

%------------%
\begin{lem}  \label{lem1}
Fix $\varepsilon>0$ and $a>1$. For $n\in \D{N}$ let
\begin{equation} \label{eq.3.4b}
a_n := \prob \left( A_{r,r+n} \text{ is not $\varepsilon$-free for
all } r \in [a^n,a^{n+1}-n] \right).
\end{equation}
Then $\Sigma_n a_n < \infty$.
\end{lem}
%------------%

%-----------%
\begin{proof}
Choose $\eta>0$ such that $a(1-\eta)>1$. Using uniform
discreteness of $\Sigma$ we get that for all $n\in \D{N}$ and
$r\ge 1$,
\begin{equation} \label{eq.3.5}
\card (A_{r,r+n} \cap \Sigma) \le Cnr^{d-1},
\end{equation}
where $C$ depends on $d$ and $r_{\Sigma}$. Here $\card A$ is the
cardinality of a set $A$. With $C$ from (\ref{eq.3.5}) choose
$\delta \in (0, \eta/(Ca^{d-1}))$.

By (\ref{eq.3.3}), $p_i(\varepsilon) \le \delta |i|^{-(d-1)}$ for
$i$ sufficiently large. Thus, for sufficiently large $n$ and each
$r\in [a^n,a^{n+1}-n]$,
\begin{eqnarray} \label{eq.3.6}
\prob (A_{r,r+n} \text{ is $\varepsilon$-free }) & = & \prod_{i\in
A_{r,r+n} \cap \Sigma} (1-p_i(\varepsilon)) \nonumber \\
& \ge & (1-\delta |i|^{-(d-1)})^{\card (A_{r,r+n} \cap \Sigma)}
\nonumber \\
& \ge & (1-\delta a^{-n(d-1)})^{Cna^{(n+1)(d-1)}} \nonumber \\
& \ge & (1-C\delta a^{d-1})^n \quad \ge \quad (1-\eta)^n.
\end{eqnarray}

$A_{a^n,a^{n+1}}$ contains at least $\frac{1}{n}(a^{n+1}-a^n)-1$
disjoint annuli $A_j := A_{r_j,r_j+n}$ of width $n$. Thus, using
independence and (\ref{eq.3.6}),
\begin{eqnarray} \label{eq.3.7}
a_n & \le & \prob (\text{no $A_j$ is $\varepsilon$-free})
\nonumber \\
& = & \prod_j \prob (A_j \text{ is not $\varepsilon$-free })
\nonumber \\
& \le & (1-(1-\eta)^n)^{n^{-1}(a^{n+1}-a^n)-1} \nonumber \\
& \le & e^{-(1-\eta)^n (n^{-1}a^n(a-1)-1)}.
\end{eqnarray}
As $(1-\eta)a>1$, the $a_n$ are summable.
\end{proof}
%-----------%

By the Borel-Cantelli lemma we conclude
$\prob(\Omega_{\varepsilon,a})=1$, where
\begin{eqnarray} \label{eq.3.7a}
\Omega_{\varepsilon,a} & := & \big\{ \omega\in \Sigma: \text{ For
each sufficiently large $n$ the annulus } A_{a^n,a^{n+1}}
\nonumber \\
& & \text{ contains a sub-annulus $A_{r_n,r_n+n}$ which is
$\varepsilon$-free for } \omega \big\}.
\end{eqnarray}
Therefore
\begin{equation} \label{eq.3.7b}
\Omega_{\varepsilon} = \bigcap_{\ell\in \D{N}}
\Omega_{\varepsilon,1+1/\ell}
\end{equation}
also has full measure.

Based on this we can now complete the

\begin{proof}[Proof of Theorem~\ref{thm.3.1}]
Fix a compact $K\subset \varrho(H_0)$. Since $\varrho(H_0)$ can be
exhausted by an increasing sequence of compact subsets, it
suffices to prove that
\begin{equation} \label{eq.3.8}
\sigma_{\text{ac}}(H(\omega))\cap K= \varnothing \text{ almost
surely.}
\end{equation}

It can be shown, using the general theory of uniformly local $L^p$
potentials, e.g.\ \cite{RSIV}, that there is an $\varepsilon'>0$
such that
\begin{equation} \label{eq.3.9}
\sigma(H_0 +V) \cap K = \varnothing,
\end{equation}
for each $V$ with $\|V\|_{p,unif} \le \varepsilon'$. Thus, by the
properties of $\Sigma$ and $f_i$, there is an $\varepsilon>0$ such
that
\begin{equation} \label{eq.3.10}
\sigma(H_0 + \sum_{i\in \Sigma} \delta_i f_i(x-i)) \cap K =
\varnothing
\end{equation}
if $\abs{\delta_i} \le \varepsilon$ for all $i\in \Sigma$.

Fix this $\varepsilon>0$ and let $\Omega_{\varepsilon}$ be the
full measure set found above. For given $\omega \in
\Omega_{\varepsilon}$ let $\tilde{\omega}_i := \min \{\omega_i,
\varepsilon\}$, $i\in \Sigma$, and
\[
V_2(x) := \sum_{i\in \Sigma} \tilde{\omega}_i f_i(x-i).
\]
By (\ref{eq.3.10}) we have $\sigma(H_0+V_2) \cap K = \varnothing$.
Thus, in order to apply Theorem~\ref{thm.ac} and conclude that
(\ref{eq.3.8}), it suffices to find for every $\gamma>0$ a total
decomposition $(S_n^{(\gamma)})$ of $\{V_{\omega} \not= V_2\}$
which satisfies (\ref{eq.2.0}) and (\ref{eq.2.1}).

For given $\gamma>0$ choose an integer $\ell>2(d-1)/\gamma$. This
implies $(d-1)\log a <\gamma/2$, where $a:= 1 +1/\ell$. As $\omega
\in \Omega_{\varepsilon,a}$, for each sufficiently large $n$ the
annulus $A_{a^n,a^{n+1}}$ contains an $\varepsilon$-free annulus
$A_{r_n,r_n+n}$.

Choose $S_n^{(\gamma)} := \partial B(0,r_n +\frac{n}{2})$. Then
\[
\delta_n^{(\gamma)} := \dist (\{ V_{\omega} \not= V_2\},
S_n^{(\gamma)}) \ge \frac{n}{2} -\rho
\]
since $A_{r_n,r_n+n}$ is $\varepsilon$-free (recall that
$\supp\,f_k \subset B(0,\rho)$). Thus $\delta_n^{(\gamma)}
\to\infty$. Also using that $\sigma(S_n^{(\gamma)}) \le
Ca^{n(d-1)}$, we conclude
\[
\sum_n \sigma(S_n^{(\gamma)}) \ee^{-\gamma \delta_n^{(\gamma)}}
\le C \ee^{\gamma\rho} \sum_n \ee^{n((d-1)\log a - \gamma/2)} <
\infty.
\]
This proves part (a) of Theorem~\ref{thm.3.1}.

In order to apply Theorem~\ref{thm.2.2} to \emph{prove part (b)}
we slightly modify the above construction, essentially replacing
$\Sigma$ by $\Sigma \setminus \{k\}$.

Let $\Omega' := \D{R}^{\Sigma\setminus \{k\}}$ with measure
$\prob' = \otimes_{i\in \Sigma\setminus \{k\}} \mu_i$. As the
property defining $\Omega_{\varepsilon,a}$ in (\ref{eq.3.7a}) does
not depend on the value of $\omega_k$, we get that also
$\prob'(\Omega_{\varepsilon,a}') = \prob'(\Omega_{\varepsilon}') =
1$, where $\Omega_{\varepsilon,a}'$ and $\Omega_{\varepsilon}'$
are defined as in (\ref{eq.3.7a}) and (\ref{eq.3.7b}), but as
subsets of $\Omega'$.

For compact $K\subset \varrho(H_0)$ choose $\varepsilon>0$ as in
the proof of part (a). For $\omega' \in \Omega_{\varepsilon}'$ let
$\tilde{\omega}_i' := \min \{\omega_i', \varepsilon\}$ ($i\in
\Sigma\setminus \{k\}$). Also let $V_{\omega'}(x) = \sum_{i\in
\Sigma\setminus \{k\}} \omega_i' f_i(x-i)$ and $V_2'(x) =
\sum_{i\in \Sigma\setminus \{k\}} \tilde{\omega}_i' f_i(x-i)$. As
before, $\sigma(H_0+V_2') \cap K = \varnothing$.

For $\gamma>0$ choose $\ell>2d/\gamma$, $a=1+1/\ell$. With $r_n$
from (\ref{eq.3.7a}), let $A_n = B(0,r_n +\frac{n}{2})$ and $S_n =
\partial A_n$. This yields
\[
|A_{n+1} \setminus A_{n-1}| \le c_d a^{(n+2)d}
\]
and
\[ \delta_n' = \min \left\{ \dist (S_n, \{V_{\omega'} \not=
V_2\}), \frac{1}{2}\dist (S_n, S_{n-1}\cup S_{n+1}) \right\} \ge
\frac{n}{2}-\rho.
\]

The choice of $a$ guarantees that $\sum_n |A_{n+1}\setminus
A_{n-1}| e^{-\gamma \delta_n'} < \infty$. By Theorem~\ref{thm.2.2}
this proves the existence of a measurable subset $M_{0,\omega'}
\subset \D{R}$ with $|\D{R} \setminus M_{0,\omega'}| =0$ and such
that
\[
\sigma_c(H(\lambda,\omega') \cap K) \subset
\sigma_c(H(\lambda,\omega') \cap \rho(H_0+V_2')) = \varnothing
\]
for all $\lambda \in M_{0,\omega'}$, where $H(\lambda,\omega') =
H_0 + \lambda f_k(x-k) + V_{\omega'}(x)$.

As $\mu_k(M_{0,\omega'}) \ge (\mu_k)_\textup{ac}(M_{0,\omega'}) =
(\mu_k)_\textup{ac}(\D{R}) = m_k$ it follows by Fubini that
$\prob \{\omega\in \Omega: \sigma_c(H(\omega))\cap K = \varnothing
\} \ge m_k$. Since this bound is independent of $K$ and we can
exhaust $\rho(H_0)$ by an increasing sequence $K_n$ we arrive at
the assertion. This completes the proof of Theorem~\ref{thm.3.1}.
\end{proof}

\begin{rem*} While the ``volume'' term $|A_{n+1}\setminus
A_{n-1}|$ in (\ref{eq.2.2}) has to be considered larger than the
``surface'' term $\sigma(S_n)$ in (\ref{eq.2.1}), this did not
make a significant difference in the above proof. The same total
decomposition $S_n$ can be used to prove absence of absolutely
continuous spectrum and absence of continuous spectrum. The
difference will become more significant for the quasi-1D surfaces
considered in the next section.
\end{rem*}

The reader will have noticed that the choice of a polynomial
bound in the assertion is somewhat arbitrary. In fact, what we use is
a Metatheorem of the form that the almost sure appearance of $\varepsilon$-free
annular regions allows one to exclude absolutely continuous spectrum
outside the unperturbed $\varrho(H_0)$. If some more regularity
holds for one of
the coupling constants, then one can even conclude that the spectrum
is pure point outside $\varrho(H_0)$ with positive probability.

Of course, the appearance of suitable geometries might also be forced
by the distribution of $\Sigma$ in space: think of $\Sigma$ that is sparse near
infinity.

%-----------%
\begin{rems*}
(1) Note that in (a) of Theorem \ref{thm.3.1} we can deal with
indefinite single site potentials and quite arbitrary single site
distributions.

(2)
Of course, the assumption $\supp\mu_j\subset\croch{0,1}$
and $0\in\supp\mu_j$ is just
to normalize things. For our methods to apply the random potentials
have to obey some uniform
bounds.

(3)
In the special case of model II considered by Hundertmark and
Kirsch our result is stronger
for $d=1$, and weaker for $d\geq 2$. However,
since they announce a proof by
multiscale analysis, it is not obvious how they want to exclude
continuous spectrum for \emph{all}
negative energies. Typically, multiscale techniques only work near
band edges. Most probably, they
implicitly have a disorder or large coupling assumption in mind.

Let us stress that our methods of proof work for \emph{all}
energies outside $\varrho(H_0)$.

(4)
In the special case of model I considered in \cite{HuKi}
the negative part of the
spectrum is purely discrete and this cannot support continuous
components. Out more general
results, however, apply in situations where the spectrum outside
$\varrho(H_0)$ fills an interval in the negative reals.
\end{rems*}
%-----------%

Using the ``Almost surely free lunch Theorem'' from \cite{HuKi}
we get the following result for $V_0=0$:

%-----------%
\begin{thm}                                    \label{thm.3.2}
Let $\mu_k$, $f_k$, $V_0$ be as above, $V_0=0$ and
assume that, additionally, the $\norm{f_k}_{\infty}$ are uniformly bounded
and that the
second moments of the $\eta_k$ obey
\[
\expect(\eta_k^2)=\int_0^1x^2\dd\mu_k\leq C\abs{k}^{-\beta}
\]
for some $\beta>2$.
Then
\[
\sigma_{\textup{ac}}(H(\omega))\supset [0,\infty)\
\prob\text{-a.s.}
\]
\end{thm}
%-----------%

%-----------%
\begin{proof}
The assumptions clearly make sure that
\[
W(x):=
\expect(V_{\omega}(x)^2)^{\frac{1}{2}}\leq C(1+\abs{x})^{-(1+\varepsilon)}
\]
so that we can apply Theorem~2.4 from \cite{HuKi} to see that
Cook's criterion is applicable for
$\prob\text{-a.e.}\;\omega\in\Omega$.
\end{proof}
%-----------%
For general $V_0$ the corresponding ist probably false. It should be true for
certain periodic potentials, see \cite{by,ger,Y}.

%-------------------------------------------------------------%
\section{Quasi-1D surfaces}                 \label{s4}

In Section~\ref{s3} sparseness of the potential $V_{\omega}$ in
(A$_4$) resulted from an assumption on decaying randomness,
e.g.~(\ref{eq.3.3}). In the present section we will modify our
methods and results for the case where sparseness of $V_{\omega}$
arises directly through sparseness of the deterministic set
$\Sigma$. By this we mean situations where $\Sigma$ does not have
positive $d$-dimensional density in $\D{R}^d$, i.e. $\# (\Sigma
\cap B(0,R)) = o(R^d)$ as $R\to\infty$. A special case would be an
$m$-dimensional sublattice, e.g.\ $\Sigma = \D{Z}^m \times \{0\}
\subset \D{R}^m \times \D{R}^{d-m}$, $0<m<d$, in which case
$V_{\omega}$ would model a random surface potential. Our most
interesting result holds for $m=1$, where our methods cover the
following more general situation:

%-----------%
\begin{Def*}
A uniformly discrete subset $\Sigma$ of $\D{R}^d$ is called
\emph{quasi-one-dimensional} (quasi-1D) if there exists $C<\infty$
such that
\begin{equation}   \label{eq.4.1}
\# (\Sigma \cap A_{R,R+1}) \le C
\end{equation}
for all $R\ge 0$.
\end{Def*}
%-----------%

%-----------%
\begin{thm}  \label{thm.4.1}
Let $H(\omega) = H_0 + V_{\omega}$ satisfy (A$_1$) to (A$_4$). In
addition, assume that $\Sigma$ is quasi-1D and that
\begin{equation}   \label{eq.4.2}
\sup_{i\in\Sigma} p_i(\varepsilon) < 1
\end{equation}
for every $\varepsilon>0$. Then $\sigma_{\text{ac}}(H(\omega))
\cap \varrho(H_0) = \varnothing$ almost surely.
\end{thm}
%-----------%

If $\Sigma$ is quasi-1D, then by Theorem~\ref{thm.4.1}, no spatial
decay in the randomness of the $\eta_i$ is required to conclude
absence of absolutely continuous spectrum in gaps of
$\sigma(H_0)$. For example, (\ref{eq.4.2}) is satisfied for
independent, identically distributed random variables $\eta_i$
such that $0\in \supp \mu$ for their common distribution $\mu$. In
particular, as every uniformly discrete $\Sigma \subset \D{R}$ is
quasi-1D, this strengthens Theorem~\ref{thm.3.1}(a) in the case
$d=1$, which would require $p_i(\varepsilon) = o(1)$ as
$k\to\infty$. Of course, in the case $d=1$ our result is hardly
new as (essentially) much stronger results are known for
one-dimensional random potentials.

More interesting is the case $d>1$, where special cases of
quasi-1D sets include discrete tubes of the form $\Sigma = \D{Z}
\times S$, with $S$ a bounded subset of $\D{Z}^{d-1}$.
Theorem~\ref{thm.4.1} shows the absence of absolute continuity in
the ``surface spectrum'' generated by the random (1D) surface
potential $V(\omega)$. Also, within certain limitations, we can
allow for curvature in the tubes $\Sigma$, thus covering rather
general ``random sausages''.

One can find quite a number of results concerning ``corrugated'' or
``random'' surfaces. Most are concerned with discrete models; see
\cite{BS:98, c1, JL, JL2, JM:98, JM:99a, JM:99b, JMP}, and
\cite{bs,HuKi} for continuoum variants. Most reminiscent of what we
have here, in \cite{JM:99a} the authors present a result
stating that the spectrum induced by a oonedimensional surface
in discrete twodimensionalö space is almost surely pure point
outside the spectrum $[-4,4]$ of the unperturbed operator.
The proof, however, is pretty much
involved and not all cases are worked out in detail.

%------------%
\begin{proof}
We start with a modification of Lemma~\ref{lem1}.

%------------%
\begin{lem}  \label{lem2}
Fix $\varepsilon>0$. Let $\delta = \sup_i p_i(\varepsilon)<1$, $C$
as in (\ref{eq.4.1}) and $a>\frac{1}{(1-\delta)^C}$. Then $a_n$,
as defined in (\ref{eq.3.4b}), is summable.
\end{lem}
%------------%

%------------%
\begin{proof}[Proof of Lemma]
This follows with the same argument as in the proof of
Lemma~\ref{lem1}, using that now $\prob (A_{r,r+n} \text{ is
$\varepsilon$-free }) \ge (1-\delta)^{Cn}$. Thus the set
$\Omega_{\varepsilon,a}$, defined as in (\ref{eq.3.7a}), has full
$\prob$-measure.
\end{proof}
%------------%

Fix $K\subset \varrho(H_0)$ compact and argue as in the proof of
Theorem~\ref{thm.3.1} to find $\varepsilon>0$ such that
$\sigma(H_0+V_2) \cap K = \varnothing$, where $V_2(x) = \sum_{i\in
\Sigma} \tilde{\omega}_i f_i(x-i)$, $\tilde{\omega}_i = \min
\{\omega_i, \varepsilon\}$. Choose $a>1$ as in Lemma~\ref{lem2}
and $\omega \in \Omega_{\varepsilon,a}$, i.e.\ $A_{a^n,a^{n+1}}$
contains $\varepsilon$-free $A_{r_n,r_n+n}$ for all sufficiently
large $n$.

As before, the spheres $S_n = \partial B(0,r_n+\frac{n}{2})$ give
a total decomposition with $\dist (\{V_{\omega} \not= V_2\}, S_n)
\ge \frac{n}{2}-\rho$. But, as Lemma~\ref{lem2} prevents us from
choosing $a$ arbitrarily close to $1$, this will not yield
convergence of (\ref{eq.2.1}) for \emph{all} $\gamma>0$. We will
therefore refine our construction by splitting the $S_n$ in two
parts. One part is a union of spherical caps for which, due to
points of $\Sigma$ close to $A_{r_n,r_n+n}$, the distance
$\frac{n}{2}-\rho$ from $\{V_{\omega} \not= V_2\}$ can't be
improved. The second part (the remaining ``swiss cheese'') has
much bigger distance to $\{V_{\omega}\not= V_2\}$ and, due to the
sparseness of $\Sigma$, contains most of $S_n$. The details of
this construction are as follows:

Fix $\alpha>1$. Let
\begin{equation} \label{eq.4.3}
P_n := (A_{r_n-n^{\alpha},r_n} \cup A_{r_n+n,r_n+n+n^{\alpha}})
\cap \Sigma
\end{equation}
be the points of $\Sigma$ in the $n^{\alpha}$-neighborhood of
$A_{r_n,r_n+n}$ (but outside $A_{r_n,r_n+n}$). For each $j\in P_n$
define the spherical cap
\begin{equation} \label{eq.4.4}
S_{n,j} := S_n \cap B((r_n+\frac{n}{2}) \frac{j}{|j|},
n^{\alpha}).
\end{equation}

%-------------%
\begin{figure}
        \includegraphics[width=12cm]{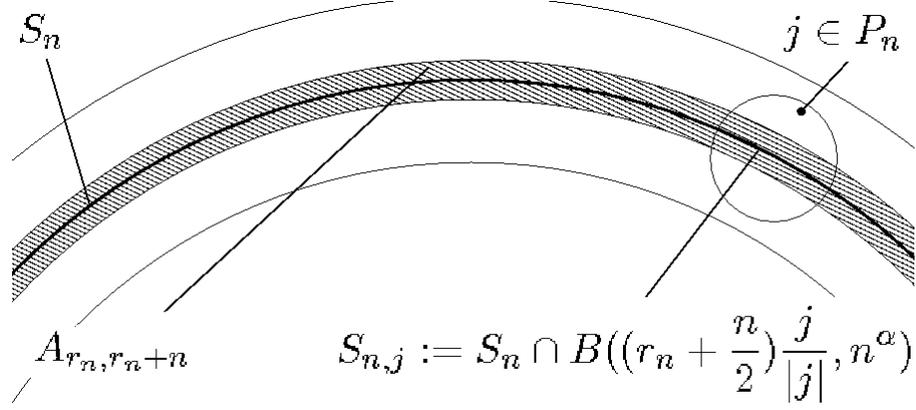}
        \caption{The geometry in the proof of Lemma 4.2: the bold face
line shows a part of $S_n$, the shaded region is $A_{r_n,r_n+n}$, the point in
the small circle a $j\in P_n$ and the small circle the boundary of
$B((r_n+\frac{n}{2})\frac{j}{|j|},
n^{\alpha})$.}
        \label{annulus}
\end{figure}
%-------------%

Also let
\[
S_n' := \overline{S_n \setminus \bigcup_j S_{n,j}}.
\]
Since $S_n' \cup \bigcup_j S_{n,j} = S_n$, we have that
\begin{equation}  \label{eq.4.5}
\{S_{n,j}: n\in \D{N}, j\in P_n\} \cup \{S_n': n\in \D{N} \}
\end{equation}
is a total decomposition of $\D{R}^d$. As above, since
$A_{r_n,r_n+n}$ is $\varepsilon$-free,
\begin{equation}  \label{eq.4.6}
\delta_{n,j} := \dist (\{V_{\omega} \not= V_2\}, S_{n,j}) \ge
\frac{n}{2}-\rho.
\end{equation}
If $x\in S_n'$ and $j\in \Sigma \cap (A_{r_n,r_n+n})^c$, then, by
elementary geometric considerations, $\dist(x,j) \ge n^{\alpha}$
for sufficiently large $n$. Using this and again that
$A_{r_n,r_n+n}$ is $\varepsilon$-free, we find
\begin{equation} \label{eq.4.7}
\delta_n' := \dist (\{V_{\omega} \not= V_2\}, S_n') \ge
n^{\alpha}-\rho.
\end{equation}
From the simple volume bound on the generalized surface area one
gets
\begin{equation} \label{eq.4.8}
\sigma(S_{n,j}) \le Cn^{d\alpha},
\end{equation}
\begin{equation} \label{eq.4.9}
\sigma(S_n') \le Ca^{dn}.
\end{equation}
Checking (\ref{eq.2.1}) for the partition (\ref{eq.4.5}) amounts
to proving that
\begin{equation} \label{eq.4.10}
\sum_n \sigma(S_n') e^{-\gamma \delta_n'} < \infty
\end{equation}
and of
\begin{equation} \label{eq.4.11}
\sum_n \sum_{j\in P_n} \sigma(S_{n,j}) e^{-\gamma \delta_{n,j}} <
\infty
\end{equation}
for each $\gamma>0$. (\ref{eq.4.10}) follows from (\ref{eq.4.7})
and (\ref{eq.4.9}) since $\alpha>1$. (\ref{eq.4.11}) follows from
(\ref{eq.4.6}) and (\ref{eq.4.8}), noting that $\# P_n \le
2n^{\alpha}+2$ since $\Sigma$ is quasi-1D. From
Theorem~\ref{thm.ac} we conclude $\sigma_{\text{ac}}(H(\omega))
\cap K \subset \sigma_{\text{ac}}(H(\omega)) \cap \varrho(H_0+V_2)
= \varnothing$.
\end{proof}
%---------%

%------------%
\begin{rem*}
It is possible to prove Theorem~\ref{thm.4.1} under a slightly
weaker assumption on the set $\Sigma$, namely that there exists
$C<\infty$ such that
\begin{equation} \label{eq.4.12}
\# (\Sigma \cap B(0,R)) \le CR
\end{equation}
for all $R\ge 1$. (\ref{eq.4.12}) is weaker than (\ref{eq.4.1}) in
that it allows the number of points in $\Sigma \cap A_{R,R+1}$ to
be unbounded with respect to $R$. (\ref{eq.4.12}) is also somewhat
more natural as it doesn't depend on the norm used to define
$B(0,R)$ nor on the choice of the center of the ball.

A simple counting argument shows that, under the assumption
(\ref{eq.4.12}), for each annulus of the form $A_{a^n,a^{n+1}}$
most sub-annuli $A_{R,R+n}$ satisfy a bound $\# (\Sigma \cap
A_{R,R+n}) \le Cn$. Here ``most'' means a non-vanishing fraction.
One finds sufficiently many disjoint such annuli to construct
$\varepsilon$-free regions as before. Moreover, by an additional
counting argument, one argues that most of these annuli do not
have more than $C'n^{\alpha}$ points of $\Sigma$ in their
$n^{\alpha}$-neighborhoods. Based on this one can construct a
partition $\{S_n', S_{n,j}\}$ as above and carry through the
proof. We skip the somewhat tedious details of this
generalization.
\end{rem*}
%------------%

We are not able to prove a result like Theorem~\ref{thm.3.1}(b),
i.e.\ absence of continuous spectrum in $\varrho(H_0)$ with
positive probability, under the assumptions of
Theorem~\ref{thm.4.1} (plus (A$_5$)). For the partition $S_n =
\partial A_n$, $A_n = B(0,r_n+\frac{n}{2})$ the volumes
$|A_{n+1}\setminus A_{n-1}|$ grow too fast to get validity of
(\ref{eq.2.2}) for all $\gamma>0$. A trick like the introduction
of $\{S_n', S_{n,j}\}$ as above is not applicable here since in
Theorem~\ref{thm.2.2} the $S_n$ need to arise as boundaries of a
growing sequence $A_n$.

However, if one replaces (\ref{eq.4.2}) by $p_i(\varepsilon) =
o(1)$ as $|i|\to\infty$ for all $\varepsilon>0$, then
Lemma~\ref{lem2} will hold for any $a>1$, which allows for an
application of Theorem~\ref{thm.2.2} with a $\gamma$-dependent
choice of the $S_n$, as in the proof of Theorem~\ref{thm.3.1}(b).
Sparseness of the random potential is achieved here through a
combination of sparseness of $\Sigma$ and decaying randomness
$p_i(\varepsilon) = o(1)$, as opposed to Theorem~\ref{thm.3.1},
where sparseness follows exclusively from stronger decay
$p_i(\varepsilon) = o(|i|^{-(d-1)})$.

In fact, the correlation between the degree of sparseness of
$\Sigma$ and the rate of decay of $p_i(\varepsilon)$ can be made
more specific. For this, call a uniformly discrete set $\Sigma
\subset \D{R}^d$ \emph{quasi-$m$-dimensional} ($1\le m \le d$, not
necessarily integer) if for some $C<\infty$ and all $R\ge 0$,
\begin{equation} \label{eq.4.13}
\# (\Sigma \cap A_{R,R+1}) \le CR^{m-1}.
\end{equation}

Then the following result is found with the same methods as above:

%-----------%
\begin{thm} \label{thm.4.3}
Let $H(\omega)$ satisfy (A$_1$) to (A$_4$), $\Sigma$ be
quasi-$m$-dimensional and, for all $\varepsilon>0$,
\begin{equation} \label{eq.4.14}
p_i(\varepsilon) = o(|i|^{-(m-1)}) \text{ as } |i|\to\infty,
\end{equation}
then $\sigma_{\text{ac}}(H(\omega)) \cap \varrho(H_0) =
\varnothing$ almost surely.

If, moreover, (A$_5$) holds, then $\prob
\{\Sigma_{\text{c}}(H(\omega)) \cap \varrho(H_0)\} \ge m_k$.
\end{thm}
%-----------%

%-------------------------------------------------------------%
\section{Concluding remarks}                 \label{s5}

Among the known results for discrete surface models, the one most
closely related to Theorem~\ref{thm.4.1} above is the result of
Jak\v{s}i\'c and Molchanov \cite{JM:99a}. They consider the
discrete Laplacian on $\D{Z} \times \D{Z}_+$ with random boundary
condition $\psi(n,-1) = V_{\omega}(n) \psi(n,0)$, where the
$V_{\omega}(n)$ are i.i.d.\ random variables. They show that the
spectrum outside $[-4,4]$, i.e.\ outside the spectrum of the
two-dimensional discrete Laplacian, is almost surely pure point.
This is stronger than our continuum analogue in the sense that we
can only prove absence of absolute continuity outside the spectrum
of the deterministic background operator $H_0$.

The proof in \cite{JM:99a} requires a technical tour de force. The
two-dimensional problem can be reduced to a one-dimensional
problem with long range interactions. Anderson localization for
the latter has been proven in \cite{JM:99c} with methods based on
an approach developed in \cite{KMP} (which is also behind
Theorem~\ref{thm.2.2} above). The one-dimensional problem depends
nonlinearly on the spectral parameter, a difficulty which is
resolved by adapting some ideas from the Aizenman-Molchanov
fractional moment method \cite{AM}.

Our methods are comparatively soft. In particular, they work
directly in the multi-dimensional PDE setting and do not require a
reduction to $d=1$. One-dimensionality of the random surface only
enters through its probabilistic consequences (Lemma~\ref{lem2})
for the frequency of the appearance of $\varepsilon$-free regions,
which constitute the ``potential barriers'' required in
Theorem~\ref{thm.ac}.

This makes our methods very flexible. In addition to the extension
to continuum models, they allow for rather general quasi-1D
surfaces (e.g.\ curved tubes, unions of tubes), work in arbitrary
dimension $d$ and allow for the presence of an additional
deterministic background potential $V_0$. It is possible to adapt
our methods to lattice operators and prove absence of absolutely
continuous spectrum outside the spectrum of the discrete Laplacian
for much more general geometries than the half-plane considered in
\cite{JM:99a}.

Also, our methods can easily be adjusted to work for operators of
the type (\ref{eq.3.2}) on $L^2(\Omega)$, $\Omega \not= \D{R}^d$.
For example, for $H(\omega) = -\Delta +V_{\omega}$ in $L^2((0,a)
\times \D{R}^{d-1})$ with Dirichlet boundary conditions and
$V_{\omega}$ given through (A$_2$) to (A$_4$) with i.i.d.\
coupling constants $\omega_i$, we would get that
$\sigma_{\text{ac}}(H(\omega)) \cap (-\infty,0) = \varnothing$
almost surely. Of course, for this physically one-dimensional
operator (with no bulk space), one would expect the much stronger
result that $\sigma_c(H(\omega)) = \varnothing$. But the
corresponding result for discrete strips, e.g.\ \cite{KLS}, does
not seem to extend easily to the continuum.

Finally, we mention that Hundertmark and Kirsch \cite{HuKi}
announce some results on pure point spectrum for continuum models
similar to the ones studied here. They will use suitable
adaptations of multiscale analysis to show that the negative
spectrum of $-\Delta +V_{\omega}$ is almost surely pure point.
Here $V_{\omega}$ is either of the type of Model~2 above or a
random potential at the surface of a half space Schr\"odinger
operator. In situations where the multiscale analysis can be
carried out, their results should be stronger than ours.

%-------------------------------------------------------------%
\subsection*{Acknowledgement}
Our collaboration has been supported by the University
Paris 7 Denis Diderot where part of this work was done, by the DFG
in the priority program ``Interacting stochastic systems of high
complexity'' and through the SFB 393, as well as through US-NSF
grant no.\ DMS-0245210.

%-------------------------------------------------------------%

%-------------------------------------------------------------%
\end{document}